\documentclass[aps,preprint,preprintnumbers,nofootinbib,showpacs]{revtex4}
\usepackage{graphicx,color,amsmath}
\newcommand{\lsim}{\mathrel{\mathop{\kern 0pt \rlap
  {\raise.2ex\hbox{$<$}}}
  \lower.9ex\hbox{\kern-.190em $\sim$}}}
\newcommand{\gsim}{\mathrel{\mathop{\kern 0pt \rlap
  {\raise.2ex\hbox{$>$}}}
  \lower.9ex\hbox{\kern-.190em $\sim$}}}
\newcommand{\beq}     {\begin{equation}}
\newcommand{\eeq}     {\end{equation}}
\newcommand{\bea}     {\begin{eqnarray}}
\newcommand{\eea}     {\end{eqnarray}}
\newcommand{\no}     {\nonumber}
\newcommand{\neu}     {\widetilde{\chi}^0}

\newcommand{\hs}      {\hat{s}}
\newcommand{\lm}      {\lambda}

\newcommand{\es}      {\epsilon}
\newcommand{\gm}      {\gamma}
\newcommand{\Gm}      {\Gamma}

\newcommand{\M}       {{\mathcal M}}

\newcommand{\imag}{\Im {\rm m}}

\newcommand{\real}{\Re {\rm e}}
\def\lsim{\ ^<\llap{$_\sim$}\ }
\def\gsim{\ ^>\llap{$_\sim$}\ }
\def\r2{\sqrt 2}
\def\chai{\widetilde{\chi}^\pm_i}
\def\cha{\widetilde{\chi}^\pm}

\def\cham{\widetilde{\chi}^-}
\def\chap{\widetilde{\chi}^+}
\begin{document}
\title{Hadronic production and decays of Charginos and Neutralinos
\\
in Split Supersymmetry}
\author{Kingman Cheung}
\affiliation{
Department of Physics and NCTS, National Tsing Hua University,
Hsinchu, Taiwan, R.O.C.
}
\author{Jeonghyeon Song}
\affiliation{
Department of Physics, Konkuk University, Seoul 143-701, Korea}
\date{\today}

\begin{abstract}
In the scenario recently proposed by Arkani-Hamed and Dimopoulos, the
supersymmetric scalar particles are all very heavy, at least of the
order of $10^9$ GeV but the gauginos, higgsino, and one of the CP-even
neutral Higgs bosons remain light under a TeV.  In addition to gluino
production, the first SUSY signature would be pair production of
neutralinos and/or charginos.  Here we study, with explicit CP
violation, the production of neutralinos and charginos at hadronic
colliders as well as the possible decay modes. We notice that the
branching ratio of the neutralino radiative decay can be sizable
unlike the case with light sfermions.  In particular, the decay of
$\widetilde{\chi}_3^0 \to \widetilde{\chi}_1^0 \gamma$
has a branching ratio of the order $O(1)$ percent.
In colliders, it would give rise to signatures of an isolated
single photon plus missing energy.
\end{abstract}
\preprint{KIAS 05036}
\maketitle

\section{Introduction}
Supersymmetry (SUSY) is one of the most elegant solutions, if not
the best, to the gauge hierarchy problem. It also provides a
dynamical mechanism for electroweak symmetry breaking, as well as a
viable candidate for dark matter (DM).
Conventional wisdom tells us that
SUSY must exist at the
TeV scale, otherwise the fine tuning problem returns.
Unfortunately these weak scale SUSY models,
represented by the minimal supersymmetric standard model (MSSM),
also have a number of difficulties
such as SUSY flavor and CP problems, and the light Higgs boson.
For the past two decades,
particle theorists have made every effort to rectify the
problems.
Recently, Arkani-Hamed and
Dimopoulos\,\cite{arkani} adopted a rather radical approach to SUSY:
they discarded the hierarchy problem and accepted the fine-tuning
solution to the Higgs boson mass.
They argued that much more serious fine-tuning
is required for the observed small cosmological constant.
If the cosmological constant problem is to be explained by
the anthropic principle\,\cite{anthropic},
an enormous number of metastable vacua,
usually called the vast landscape of string theory vacua,
are essential.
Fine-tuning in the Higgs boson mass is more natural in this anthropic
landscape\,\cite{landscape}.

Once we accept this proposal the finely-tuned Higgs boson mass is not a
problem anymore.  More concerned issue is to find a consistent set of
parameters satisfying the following observations:
(i) the refined result of the
DM density to be $\Omega_{\rm DM} h^2 = 0.094 - 0.129$ ($2\sigma$ range)
by
Wilkenson Microwave Anisotropy Probe (WMAP)\,\cite{wmap};
(ii) sub-eV neutrino masses; and
(iii) cosmological constant.
The last one is accepted
as an extremely fine-tuned principle.
The second observation requires
heavy right-handed neutrinos of a mass scale of $10^{11-13}$ GeV;
it does not have appreciable effects on electroweak scale physics.  The first
observation, on the other hand, requires a weakly interacting particle of
mass $\alt 1$ TeV, in general.  It is this requirement which affects most of
the parameter space of the split SUSY scenario\,\cite{giudice,adgr}.

The split SUSY scenario can be summarized as follows:
\begin{enumerate}
\item All scalars, except for a CP-even neutral
Higgs boson, are super heavy,
for which a common mass scale $\tilde{m} $ is assumed
around $10^{9}$ GeV to $ M_{\rm GUT}$.
Dangerous phenomena such as flavor-change neutral currents,
CP-violating processes, and large electric dipole moments
in the MSSM become safe.\footnote
{There is another source of $CP$-violation as pointed out in Ref.\,\cite{edm}.
}
\item The gaugino and the Higgsino masses (the $\mu$ parameter),
which can be much lighter than $\tilde{m}$ due to
an $R$ and a PQ symmetry, respectively,
are assumed near the TeV scale.
Light supersymmetric fermions
have additional virtue of gauge coupling unification
as in MSSM,
as well as providing a good DM candidate.
\item A light Higgs boson is very similar to the SM Higgs boson,
but could be substantially heavier than that of the
MSSM\,\cite{arkani,giudice,unif,binger,mahbubani}.
\item The DM density requires sufficient mixing in the neutralino sector,
so that the lightest neutralino can efficiently
annihilate\,\cite{aaron,stefano,giudice}.
In most cases, the $\mu$ parameter is relatively small.
\end{enumerate}

Super heavy masses of SUSY scalar particles practically
close off many neutralino annihilation channels,
resulting in smaller annihilation cross section
than that in MSSM.  Giudice and Romanino\,\cite{giudice}, Pierce\,\cite{aaron},
and Masiero et al.\,\cite{stefano}
identified three
interesting regions of the parameter space
where the split SUSY model can accommodate the
WMAP data on the DM density.
\begin{itemize}
\item[(i)] The lightest neutralino is mostly
a Bino, but with a substantial mixing with the Higgsino: $\mu$ is
comparable to $M_1$.  This is because
the Bino interacts only with the
Higgs and Higgsino when all sfermions are super heavy.
Substantial mixing can guarantee efficient annihilation
of the Bino
into Higgs or $Z$ bosons.
The WMAP data allow the gaugino mass parameters as low as
the current lower bound.
\item[(ii)] The second region
is where the LSP is mostly Higgsino with $M_{1,2} \gg \mu$.
In this case, the Higgsino LSP annihilates via
gaugino-Higgsino-Higgs and Higgsino-Higgsino-gauge couplings.
Note that the lightest chargino and the second lightest neutralino
are also predominately Higgsino since $\mu \ll M_{1,2}$,
and thus degenerate in mass with the LSP.
Efficient annihilation and co-annihilation require a rather heavy
LSP mass $\mu \sim 1-1.2$ TeV for the LSP to be the DM.
\item[(iii)] The third region is the wino LSP
in anomaly mediation and in this case
$M_2 < M_1, \mu$.  Neutral wino can annihilate efficiently via an intermediate
$W^*$, even in the absence of sfermions.  Thus, a rather heavy wino
with $M_2\simeq 2.0-2.5$ TeV is
required to account for the dark matter density.
\end{itemize}
The last two cases require a rather heavy LSP.  Therefore,
the chance of observing supersymmetric partners at colliders
diminishes as the gluino would be even heavier than 2--3 TeV,
though it is not impossible.  Thus, we concentrate on the
phenomenologically more interesting case where $M_1 \sim \mu
\sim M_{\rm EW}$.
Note that once we impose the condition of gaugino-mass unification,
$M_2 \simeq 2 M_1 $ at the weak scale.  The current chargino mass
bound is almost 104 GeV \cite{bound}.  If converting to the
bounds on $M_2$ and $\mu$, both $M_2$ and $\mu$ larger than $O(100)$ GeV
should be consistent with the bound.  We do not specifically
state the bound on $M_1$ and $\mu$, but $O(100)$ GeV should be very safe
for $M_1$ and $\mu$ with the chargino mass bound.

The detailed phenomenological study of the split SUSY scenario,
especially focused on high energy colliders,
is crucial to clarify the basic structure of the scenario.
As pointed out in Ref.\,\cite{arkani},
the first unique feature is the stable or metastable gluino.
It may give rise to stable charged tracks\,\cite{kilian,rizzo,anch} or
gluinonium\,\cite{yee} signatures.
We focus on the neutralino pair, chargino pair, and
neutralino-chargino pair production, and their decays at hadron colliders.
These production channels may be the only accessible ones if the
gluino mass is more than $3$ TeV.
There have been some studies of chargino and neutralino production
at $e^- e^+$ colliders\,\cite{zhu} and at hadron colliders\,\cite{kilian}.
Some variations on split SUSY were also studied, e.g., taking
$\mu$ to be very large\,\cite{cw}, taking gaugino masses to be very
large\,\cite{high-gau}, or even taking $\mu$ and all gaugino masses to be very
heavy\,\cite{fun}.
\footnote{The last scenario was called supersplit supersymmetry which
appeared in arXic.org on April fool's day.  It is equivalent to the
SM.}
Note that the phenomenology of split SUSY is rather similar to those of
focus-point SUSY \cite{0507032}, except for the gluino phenomenology.
There is also the PeV scale SUSY scenario \cite{jim-wells}.

At hadron colliders,
the production of $\neu_i\,\neu_j$, $\neu_j\,\chai$, and
$\widetilde{\chi}^+_i\, \widetilde{\chi}^-_j$
is mainly via the Drell-Yan-like processes
with intermediate $W^*, \gamma^*$, or $Z^*$ exchanges
as in the usual MSSM.
The $t$-channel exchanges
via $\tilde{q},\tilde{\ell}$ are virtually absent.
On the other hand, the decays of the
neutralino $\neu_j(j>1)$ and chargino would be quite different from
the MSSM case.
With super-heavy sfermions the decay of the heavier
neutralino $\widetilde{\chi}^0_j$ into a lighter neutralino
$\widetilde{\chi}^0_i$ plus $f\bar f$ is mainly via an intermediate
$Z^*$ (the Higgsino component only) instead of a $\tilde{f}$.
Another channel of importance is $\widetilde{\chi}^0_j \to\cha_i W^*
\to \cha_i f\bar{f}'$.
We also have $\widetilde{\chi}^0_j \to\widetilde{\chi}^0_i
h^{(*)} \to \widetilde{\chi}^0_i + (b\bar b \;{\rm or}\; WW^*)$.
Finally, we note that the branching ratio of the radiative decay
$\widetilde{\chi}^0_j \to
\widetilde{\chi}^0_i \gamma$ can be sizable unlike the case with
light sfermions.
Phenomenologically it leaves clean signal at the LHC,
consisting of
a high-energy single-photon plus missing energy.
Particularly in the parameter region where $|\mu|$ is comparable
to $M_1$, the radiative branching ratio would be maximal.
The chargino decays mainly through
$\widetilde{\chi}^+_j \to \widetilde{\chi}^0_i + W^* \to
\widetilde{\chi}^0_i + f \bar f'$.
By combining the phenomenological studies on neutralinos and charginos,
as well as those of gluino, one may be able to distinguish the scenarios of
split SUSY, PeV scale SUSY, or focus-point SUSY.

Improvement over previous studies\,\cite{kilian,zhu}
can be summarized as follows.
\begin{enumerate}
\item We include all decay modes of neutralinos and charginos, in
particular, the radiative decay mode.  It gives rise to a single-photon plus
missing energy signal.  Experimentally, it is very clean.
\item We include the CP-violating phases in $M_1$ and
 $\mu$
in order to examine the effect of the CP phases on the decay of neutralinos.
\item We calculate the photon transverse momentum and missing
transverse momentum distributions for the single-photon plus missing
energy events, which mainly
come from the decay of $\widetilde{\chi}^0_3 \to \widetilde{\chi}^0_1 \gamma$.
\end{enumerate}

The organization of this paper is as follows.  We describe
the general formalism and convention
in Sec. II.  We calculate the decays and production cross sections
for neutralino and chargino pairs in Sec. III and IV, respectively.
We conclude in Sec. V.

\section{General Formalism of neutralinos and charginos}
The neutralino and chargino sectors are
determined by fundamental SUSY parameters:
the U(1) and
SU(2) gaugino masses $M_1$ and $M_2$, the Higgsino mass parameter
$\mu$, and $\tan\beta=v_2/v_1$
(the ratio
between the vacuum expectation
values of the two neutral Higgs fields).
The neutralino mass matrix in the
$(\widetilde{B},\widetilde{W}^3,\widetilde{H}^0_1,\widetilde{H}^0_2)_L$ basis,
where the subscript $L$ denotes the left-handed chirality of neutralinos,
is\,\cite{Choi:neutralino}
\begin{eqnarray}
{\cal M}_N
=\left(\begin{array}{cccc}
  M_1       &      0          &  -m_Z c_\beta s_W  & m_Z s_\beta s_W \\[2mm]
   0        &     M_2         &   m_Z c_\beta c_W  & -m_Z s_\beta c_W\\[2mm]
-m_Z c_\beta s_W & m_Z c_\beta c_W &       0       &     -\mu        \\[2mm]
 m_Z s_\beta s_W &-m_Z s_\beta c_W &     -\mu      &       0
                  \end{array}\right)\
\label{eq:massmatrix}
,
\end{eqnarray}
where $s_\beta =\sin\beta$, $c_\beta=\cos\beta$ and
$s_W,c_W$ are the sine and cosine of the electroweak mixing angle
$\theta_W$.
Since ${\cal M}_N$ is symmetric, one
unitary matrix $N$ can diagonalize the $\M_N$
such that $
{\cal M}_{diag}=N^* {\cal M}_N N^{\dagger}
$. The Majorana mass eigenstates are
\beq
   \left(
           \widetilde{\chi}^0_1 ,
           \widetilde{\chi}^0_2 ,
           \widetilde{\chi}^0_3 ,
           \widetilde{\chi}^0_4
             \right)_L^T
   =  N\, \big(
           \widetilde{B} ,
           \widetilde{W}^3 ,
           \widetilde{H}^0_1 ,
           \widetilde{H}^0_2
             \big)_L^T.
\eeq
The mass eigenvalues $m_{\widetilde{\chi}^0_i}$ $(i=1,2,3,4)$ in ${\cal
M}_{diag}$ can  be chosen
positive by a suitable definition of the unitary matrix $N$.

The chargino mass matrix in the
$(\widetilde{W}^-,\widetilde{H}^-)$ basis is\,\cite{Choi:chargino}
\begin{eqnarray}
{\cal M}_C=\left(\begin{array}{cc}
                M_2                &      \sqrt{2}m_W\cos\beta  \\
             \sqrt{2}m_W\sin\beta  &             \mu
                  \end{array}\right)\,,
\label{eq:mass matrix}
\end{eqnarray}
which is diagonalized by two unitary matrices\footnote{
$U_{L}$ and $U_R$ are related to $U$ and $V$ in
Haber-Kane\,\cite{kane} notation by
$
V =U_R^*$ and $ U= U_L.
$
} through
$
U_R {\cal M}_C U_L^\dagger = \mathrm{diag}(m_{\cha_1},m_{\cha_2})$.
The unitary matrices $U_L$ and $U_R$ can be parameterized in the
following way\,\cite{Choi:chargino}:
\beq
U_L=\left(\begin{array}{cc}
             \cos\phi_L & {\rm e}^{-i\beta_L}\sin\phi_L \\
            -{\rm e}^{i\beta_L}\sin\phi_L & \cos\phi_L
             \end{array}\right),\quad
U_R=\left(\begin{array}{cc}
     {\rm e}^{i\gamma_1}\cos\phi_R & {\rm e}^{i(\gamma_1-\beta_R)}\sin\phi_R \\
    -{\rm e}^{i(\gamma_2+\beta_R)}\sin\phi_R & {\rm e}^{i\gamma_2}\cos\phi_R
             \end{array}\right)
             \,.
\eeq

In CP-violating theories, the mass parameters are
complex.
Since $M_2$
can be taken real and positive by rephasing the fields suitably,
the split SUSY scenario allows only two CP-violating phases of
$M_1$
and $\mu$:
\begin{eqnarray}
M_1=|M_1|\,\,{\rm e}^{i\Phi_1}\ \quad {\rm and} \quad
\mu=|\mu|\,\,{\rm e}^{i\Phi_\mu} \quad (0\leq \Phi_1,\Phi_\mu< 2\pi)\;.
\end{eqnarray}
The universal gaugino mass relation at the GUT scale implies at the
weak scale
\beq
|M_1| = \frac{5}{3}\tan^2\theta_W M_2 \simeq 0.502 M_2
\,.
\eeq
The five underlying SUSY parameters
$\{|M_1|, \Phi_1, |\mu|, \Phi_{\mu}; \tan\beta\}$
determine the mass spectrum and couplings of the neutralinos and charginos.

The interaction Lagrangian relevant for the production and decay
of neutralinos and charginos are expressed in 4-component form as
\begin{eqnarray}
{\cal L} &=&  e  \sum_i
 \overline{\widetilde{\chi}^-_i} \gamma^\mu  \widetilde{\chi}^-_i A_\mu
 -g_Z \sum_{\alpha,i,j} \,Q^{\rm ccZ}_{\alpha i j}\,
\overline{\widetilde{\chi}^-_i} \gamma^\mu
 P_\alpha \widetilde{\chi}^-_j Z_\mu
-g_Z \sum_{\alpha,i,j}\,Q^{\rm nnZ}_{\alpha i j}\,
\overline{\widetilde{\chi}^0_i} \gamma^\mu
P_\alpha \widetilde{\chi}^0_j Z_\mu
\nonumber \\
&&- g \sum_i \,S_i^{\rm nnh} \,\overline{\neu_i}{\neu}_i h^0
- g \sum_{i<j}\,Q^{\rm nnh}_{\alpha ij}\,\overline{\neu_i} P_\alpha \neu_{j}h^0
\nonumber \\
&&-g \sum_{\alpha,i,j} \,Q^{\rm cnW}_{\alpha i j}\,
\overline{\widetilde{\chi}^-_i}\gamma^\mu P_\alpha \neu_{j}W_\mu
+H.c. \nonumber \\
&&
-g \sum_{\alpha,i,j} \,Q^{\rm cnH}_{\alpha i j}\,
\overline{\widetilde{\chi}^-_i} P_\alpha \neu_{j}H^-
-g \sum_{\alpha,i,j} \,Q^{\rm cnG}_{\alpha i j}\,
\overline{\widetilde{\chi}^-_i} P_\alpha \neu_{j}G^-+H.c.
\,,
\label{lag}
\end{eqnarray}
where $\alpha=L,R$, $P_{R,L}=(1\pm \gamma^5)/2$,
$i,j=1,2$ for the chargino and $i,j=1,\cdots 4$ for the
neutralino.\footnote{Note that $\cham$ is considered as a particle,
contrary to the Haber and Kane notation.}
Here $H^-$ is the charged Higgs boson and $G^-$ is the charged Goldstone
boson.
In the non-linear $R_\xi$ gauge
used in the calculation of the radiative decay,
the $G^\pm$ has the mass of $m_W$, of which the contribution is
important.
The detailed expressions for various couplings $Q_{\alpha i j}$ are listed in
Appendix.

\section{Decays of neutralinos and charginos}

In general, the production cross section of gaugino-pair in split SUSY
is smaller than the case with light sfermions.  It brings a
challenge to experimental search of gauginos.
%
%
At the LHC, the gluino-pair production is dominated by the 
$gg$-initiated subprocess, which stays the same as in the MSSM. (Though 
the $q\bar q$-initiated subprocess changes, it is very small compared 
with $gg$-initiated one.)
On the other hand, the detection is very different.
Gluinos so
produced would be detected as massive stable charged
particles\,\cite{kilian,rizzo,anch} or as a gluinonium\,\cite{yee}.
Even though the detection of heavy meta-stable gluinos alone
can strongly support the split SUSY scenario,
it cannot provide any more information on the fundamental SUSY parameters
involved in the neutralino and chargino sectors.
Even with small production rates, the decays of neutralinos and charginos
in split SUSY are to be studied in detail.

In split SUSY,
the chargino decay is very simple.
It would decay into $\widetilde{\chi}^0_1 W^{(*)}$ giving rise to
a single charged lepton or two jets plus missing energy.
On the other hand, the heavier
neutralino $\widetilde{\chi}^0_j$ have a few more decay modes.
\begin{itemize}
\item $\widetilde{\chi}^0_j \to \widetilde{\chi}^0_i Z^{(*)} \to
\widetilde{\chi}^0_i (jj/\ell\bar \ell)$, giving rise to a couple of jets or
charged leptons plus missing energy.
\item
$\widetilde{\chi}^0_j \to \widetilde{\chi}^0_i h^{(*)} \to
\widetilde{\chi}^0_i (b\bar b/ W W^*)$.
If the Higgs boson is lighter than about 125 GeV, it decays dominantly into
$b\bar b$.
With $m_h \ge 130$ GeV the $WW^*$ mode
becomes important, which is possible
because the Higgs mass can be substantially larger in split SUSY
\cite{giudice,unif,binger,mahbubani}.
\item
$\widetilde{\chi}^0_j \to \widetilde{\chi}^\pm_i W^{\mp(*)} \to
\widetilde{\chi}^\pm_i f \bar f'$.
This mode happens when the heavier neutralino is heavier than the
lightest chargino, especially, in the region where the $\mu$ parameter
is smaller than $M_2$.
\item
$\widetilde{\chi}^0_j \to \widetilde{\chi}^0_i \gamma$\,\cite{haber}.
This mode goes
through the chargino-$W$ loop,
as the loops involving sfermions
are highly suppressed.  This decay mode gives a single photon and
missing energy.
\end{itemize}
The expressions for
the decay rates of first three decay modes are greatly simplified in
split SUSY.  We give the formulas for various decays
of the neutralino $\widetilde{\chi}^0_j$ into $\widetilde{\chi}^0_i$.
Note that for decays of $\widetilde{\chi}^0_{3,4}$ they may go through
$\widetilde{\chi}^0_{2}$ before they end up in $\widetilde{\chi}^0_1$.

\subsection{Two body decays of
$\widetilde{\chi}^0_j \to \widetilde{\chi}^0_i Z,~
\widetilde{\chi}^0_i h^0,~
\chai W^\mp$}
For simplicity we introduce some short-hand notation of
\begin{equation}
 \mu_{ij}= \left(\frac{m_{\widetilde{\chi}^{0(\pm)}_i}}
{m_{\widetilde{\chi}^0_j}} \right)^2\;, \qquad
 \mu_{Xj} = \left(\frac{m_{X}}{m_{\widetilde{\chi}^0_j}}\right)^2\,,
\end{equation}
where $X=Z,h^0,W^\pm$.
If $m_{\widetilde{\chi}^0_j} > m_{\widetilde{\chi}^{0(\pm)}_i} + m_X$,
$\widetilde{\chi}^0_j$
decays into a lighter neutralino or chargino associated with the $X$ boson.
The total decay rate is then
\beq
\Gamma\left(
\neu_j \to \widetilde{\chi}^{0(\pm)}_i X
\right) =
\frac{\lm^{1/2}(1,\mu_{ij},\mu_{Xj})}{16\pi m_{\neu_j}}
\,
\overline{|{\cal M}|^2}
\,,
\eeq
where $\lm(a,b,c)=a^2+b^2+c^2-2 a b -2 b c-2 ca$
and  $\overline{|{\cal M}|^2}$ is the spin-average amplitude squared.
For each decay mode
$\overline{|{\cal M}|^2}$ is
\bea
\overline{|{\cal M}|^2}(\widetilde{\chi}^0_j \to \widetilde{\chi}^0_i Z)
&=&
\frac{g_Z^2 m_{\neu_j}^2}{2}
\left[
(|Q^{\rm nnZ}_{Rij}|^2+|Q^{\rm nnZ}_{Lij}|^2)
\left\{
\frac{(1-\mu_{ij})^2}{\mu_{Zj}} + 1+\mu_{ij}-2\mu_{Zj}
\right\}
\right.
\\ \nonumber
&& ~~~~~~~~
-12 \sqrt{\mu_{ij}}\, \Re e(Q^{\rm nnZ}_{Rij} Q^{\rm nnZ*}_{L ij})
\Big],
\\
\overline{|{\cal M}|^2}
\left(
\neu_j \to \neu_i h^0
\right)
&=&
\frac{g^2 m_{\neu_j}^2}{2}
\Big[
\left(|Q^{\rm nnh}_{Rij}|^2+|Q^{\rm nnh}_{Lij}|^2\right)(1+\mu_{ij}-\mu_{hj})
\\ \nonumber
&& ~~~~~~~~
+4\sqrt{\mu_{ij}}\,
\Re e(Q^{\rm nnh}_{Rij} Q^{\rm nnh*}_{Lij})
\Big]
\,,
\\
\overline{|{\cal M}|^2}(\neu_j \to \cham_i W^+)
&=& \overline{|{\cal M}|^2}(\neu_j \to \chap_i W^-)
\\ \no
&=&\frac{g^2 m_{\neu_j}^2}{2}
\left[
(|Q_{L1j}^{\rm cnW}|^2 + |Q_{R1j}^{\rm cnW}|^2)
\left(\frac{(1-\mu_{ij})^2}{\mu_{Wj}}+1+\mu_{ij}-2\mu_{Wj}\right)
\right.
\\ \no &&~~~~~~~~
\left.
-12 \sqrt{\mu_{ij}}\,\Re e(Q_{L1j}^{\rm cnW}Q_{R1j}^{\rm cnW*})
\right]
\,.
\eea

\subsection{Three body decays of $\neu_j \to
\widetilde{\chi}^{0,\pm}_i f \bar{f}^{(\prime)}$}
If
$m_{\widetilde{\chi}^0_j}< m_{\widetilde{\chi}^{0(\pm)}_i} + m_X$,
the decay will
proceed into $\widetilde{\chi}^{0(\pm)}_i f\bar f^{(')}$
via a virtual $Z^*$, $W^*$ or $h^*$.
The differential
decay width is given by
\begin{equation}
\frac{d\Gamma}{d x_f d x_{\bar f}} =
\frac{N_C m_{\widetilde{\chi}^0_j} }{256 \pi^3}\,
\overline{|\M|^2}\,,
\end{equation}
where $N_C$ is the color factor of the fermion $f$
and the kinematic variables are defined in the rest frame of
$\neu_j$ by
\beq
x_f = \frac{2 E_f}{m_{\widetilde{\chi}^0_j}}\,, \quad
x_{\bar f} = \frac{2 E_{\bar f}} {m_{\widetilde{\chi}^0_j}}\,,
\quad
x_i = 2-x_f-x_{\bar{f}}
\,.
\eeq
The integration range of $x_f$ and $x_{\bar f}$ are,
with the definition of $\mu_{f(\bar f)}\equiv m_{f(\bar f)}^2/m_{\neu_j}^2$,
\begin{eqnarray}
2 \sqrt{\mu_{f}} \le &x_f & \le 1+\mu_{f}- \mu_{\bar f }
- \mu_{ij} - 2\sqrt{
 \mu_{\bar f } \mu_{ij} }  \;,  \label{inte} \\
 x_{\bar{f}_{(-)}}\le &x_{\bar f}& \le x_{\bar{f}_{(+)}}
\nonumber
\end{eqnarray}
where
\bea
x_{\bar{f}_{(\pm)}}=
\frac{1}{2(1-x_f +\mu_{f})} &\Big[&
(2-x_f)(1+\mu_{f} + \mu_{\bar f } - \mu_{ij} - x_f) \\ \nonumber&&\pm
\sqrt{(x_f^2 - 4 \mu_{f} )\lambda(1+\mu_{f } - x_{f}, \, \mu_{\bar {f }},\,
\mu_{ij} )} \;\Big ]
\,.
\eea
The spin-average amplitude squared for each decay mode is
\bea
\overline{|\M|^2}\!\!\!\!\!&&\!\!\!\!\!(\neu_j \to \neu_i Z^*
\to \neu_i f\bar{f})
\\ \nonumber
&=&g_z^4(|g^f_L|^2 +|g^f_R|^2) \hat{d}_Z^2
\Big[
\{
(x_f+x_{\bar{f}})(1-\mu_{ij})- x_f^2 - x^2_{\bar{f}}
\}(|Q^{\rm nnZ}_{Lij}|^2 +|Q^{\rm nnZ}_{Rij}|^2)
\\ \no &&
-4 \sqrt{\mu_{ij}} \,(1+\mu_{ij}-x_i)
\Re e (Q^{\rm nnZ}_{Lij}Q^{\rm nnZ*}_{Rij})
\Big]
\,,
\\
\overline{|\M|^2}\!\!\!\!\!&&\!\!\!\!\!(\neu_j \to \neu_i h^* \to
\neu_i b \bar{b})
\\ \no
&=&
\frac{g^4 m_b^2 \sin^2\alpha}{4 m_W^2 \cos^2\beta} \hat{d}_h^{2}
(1+\mu_{ij}-x_i- 2 \mu_b)
\left[
x_i ( |Q_{Lij}^{\rm nnh}|^2 + |Q_{Rij}^{\rm nnh}|^2)
+ 4\sqrt{\mu_{ij}}\, \Re e (Q_{Lij}^{\rm nnh}Q_{Rij}^{\rm nnh*})
\right]
\,,
\\
\overline{|\M|^2}\!\!\!\!\!&&\!\!\!\!\!(\neu_j \to \cham_i W^* \to
\cham_i f \bar{f}')
\\ \no
&=& g^4\hat{d}_W^2
\left[
x_f(1-\mu_{ij}-x_f) |Q_{Lij}^{\rm cnW}|^2
+
x_{\bar{f}}(1-\mu_{ij}-x_{\bar{f}}) |Q_{Rij}^{\rm cnW}|^2
\right.
\\ \no && \left.
- 2\sqrt{\mu_{ij}}\,
(1+\mu_i-x_i)\,
\Re e (Q_{Lij}^{\rm cnW}Q_{Rij}^{\rm cnW*})
\right]
\,,
\eea
where the propagator factor $\hat{d}_X$, with $X=Z,h^0,W$,
is
\beq
\hat{d}_X = \frac{m_{\neu_j}^2}{(p_f+p_{\bar{f}})^2-m_X^2}
=\frac{1}{1+\mu_{ij}-x_i-\mu_{Xj}}
\,.
\eeq
Here the chiral couplings of the fermion $f$ to the $Z$ boson are given
by
\beq g^f_R =
-s_W^2 Q_f,\quad g^f_L = (T^f_3)_L- s_W^2 Q_f \,,
\eeq
where $(T^f_3)_L$ is the third component of the isospin and
$Q_f$ is the electric charge.

\subsection{Radiative decay}

In split SUSY, the radiative decay of neutralino,
$\neu_j \to \neu_i \gm$,
may have a substantial branching fraction
as the other decay channels
are limited to
$\neu_j \to \neu_i Z^{(*)}$,
$\neu_j \to \neu_i h^{(*)}$,
and $\neu_j \to \widetilde{\chi}^\pm_i W^{\mp(*)}$.
As discussed in Ref.\,\cite{haber},
the radiative neutralino decay proceeds through triangle diagrams
mediated by the $f\tilde{f}$, $\cha W^\mp$, $\cha H^\mp$,
and $\cha G^\mp$ loops in the non-linear $R_\xi$ gauge
where the charged Goldstone boson $G^\pm$ has the same mass as $m_W$.
In split SUSY, only $W^\pm$- and $G^\pm$-mediated diagrams contribute.
We expand the results of Ref.\,\cite{haber}
into the CP-violating case,
and examine whether CP-violating phases can enhance the radiative decay rate
of neutralinos.

For the decay of
\beq
\neu_j(p) \to \neu_i(k_1)+ \gm(k_2),
\eeq
the matrix element is, in general, given by
\beq
\mathcal{M} = \frac{1}{m_{\neu_j}}
\bar{u}(k_1)
\left(
g_v+g_a \gm_5
\right)\rlap/k_2 \rlap/\es^* u(p)
\,,
\eeq
where $\epsilon$ denotes the polarization 4-vector of the photon,
and the radiative decay width of $\neu_j$ can be easily calculated as
\beq
\Gm (\neu_j \to \neu_i \gm)
=(|g_v|^2+|g_a|^2)\frac{(m_{\neu_j}^2-m_{\neu_i}^2)^3}{8\pi m_{\neu_j}^5}
\,.
\eeq

In the notation of Ref.\,\cite{haber},
the coupling of the incoming neutralino $\neu_j$ to the particles
in the loop is denoted by
\beq
 F^X = f^X_L P_L + f^X_R P_R
 \,,
\eeq
and the coupling of the outgoing neutralino $\neu_i$ by
\beq
 G^X = g^X_L P_L + g^X_R P_R
 \,,
\eeq
where $X=W^\pm, G^\pm$ and $-i g \gm^\mu (-i g)$ is omitted
for the $W^\pm(G^\pm)$ loop respectively.
Explicitly we have
\bea
f^W_\alpha &=&  Q^{\rm cnW}_{\alpha k j }, \quad
g^W_\alpha =  (Q^{\rm cnW}_{\alpha ki})^*,
\\ \no
f^G_\alpha &=& Q^{\rm cnG}_{\alpha k j }, \quad
g^G_\alpha = (Q^{\rm cnG}_{\bar\alpha ki})^*,
\eea
where $\bar\alpha=R(L)$ for $\alpha=L(R)$.
The helicity amplitude in the CP-violating cases is
\beq
\mathcal{M} =\sum_{X=W,G} \frac{1}{m_{\neu_j}}\frac{e g^2}{8\pi^2}
\bar{u}(k_1)(g_V^{X}+g_A^{X}\gm_5)\rlap/k_2
\rlap/\es^* u(p)
\,,
\eeq
where  $g_{V,A}^X$ are given by
\bea
g_V^{W} &=&i\sum_{k=1,2}
\left[
\,\imag (g_L^W f_L^W + g_R^W f_R^W  )
\left\{
m_{\neu_j}^2 (I_2-J-K)-m_{\neu_j} m_{\neu_i}(J-K)
\right\}
\right.
\\ \no
&&
\left.~~~~~~~~
+
2m_{\neu_j} m_{\cha_k}\, \imag (g_L^W f_R^W + g_R^W f_L^W)\, J
\right],
\\ \no
g_A^{W} &=&\sum_{k=1,2}
\left[
\,\real (g_L^W f_L^W  - g_R^W f_R^W  )
\left\{
m_{\neu_j}^2 (I_2-J-K)+m_{\neu_j} m_{\neu_i}(J-K)
\right\}
\right.
\\ \no
&&
\left.~~~~~~~~
+
2m_{\neu_j} m_{\cha_k} \, \real (g_L^W f_R^W - g_R^W f_L ^W )\,J
\right], \\ \no
g_V^{G} &=& \frac{i}{4} \sum_{k=1,2}
\left[
\,\imag (g_L^G f_R^G  + g_R^G f_L^G)
\left\{
m_{\neu_j}^2 (I_2-K)+m_{\neu_j} m_{\neu_i} K
\right\}
\right.
\\ \no
&&
\left.~~~~~~~~
+m_{\neu_j} m_{\cha_k}
\, \imag (g_L^G f_L^G+g_R^G f_R^G)I
\right],
\\ \no
g_A^{G} &=&-\frac{1}{4} \sum_{k=1,2}
\left[
\,\real (g_L^G f_R^G - g_R^G f_L^G  )
\left\{
m_{\neu_j}^2 (I_2-K)-m_{\neu_j} m_{\neu_i} K
\right\}
\right.
\\ \no
&&
\left.~~~~~~~~
+ m_{\neu_j} m_{\cha_k} \,
\real (g_L f_L - g_R f_R )I
\right]
\,.
\eea
The radiative decay width is
\beq
\Gm = \frac{\alpha^3}{8 \pi^2 \sin\theta_W^4}
(|g_V|^2+|g_A|^2)\frac{(m_{\neu_j}^2-m_{\neu_i}^2)^3}{m_{\neu_j}^5}
\,,
\eeq
where $g_{V,A} \equiv \sum_X g_{V,A}^X$.
The loop integrals $I,I_2,J,K$ are the same as in Ref.\,\cite{haber}.

\subsection{Numerical Results of neutralino decay rates}

Focusing on the parameter space of $|\mu|$ compatible with
$M_1$,
we consider $|\mu| \in [160,\,230]$ GeV for $M_1=200$ GeV.
With the gaugino mass unification condition of
$|M_1| \simeq 0.502 M_2
$,
the mass hierarchy among the neutralinos and charginos is then
$m_{\neu_1} < m_{\cha_1} < m_{\neu_2} < m_{\neu_3}$.

\begin{figure}[t!]
\begin{center}
 \includegraphics[scale=1.]{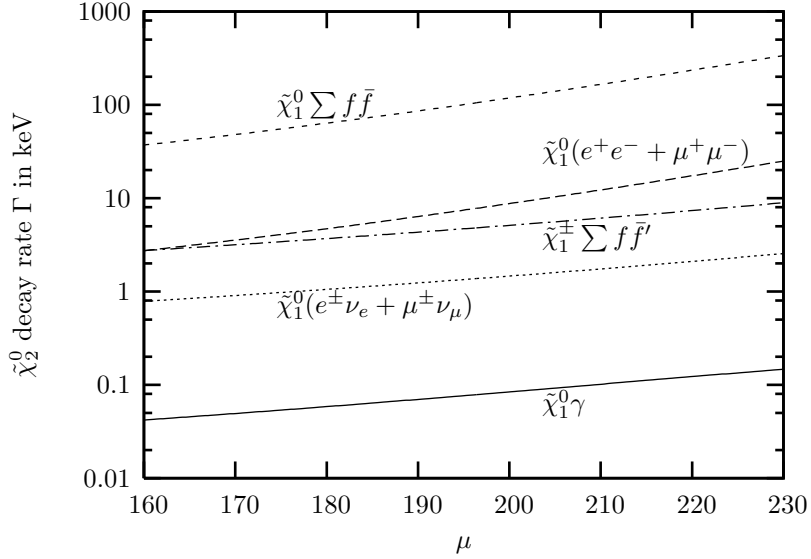}
\end{center}
  \caption{\label{fig:Gam2}
Partial decay widths of $\neu_2$
for $M_1=200$ GeV and $\tan\beta=10$.
All the CP violating phases are set to zero.
$\sum f \bar{f}^{(\prime)}$ includes all possible SM fermion pairs
except for $e^\pm$ and $\mu^\pm$ involving pairs.
}
\end{figure}

\begin{figure}[t!]
\centering
  \includegraphics[scale=1.]{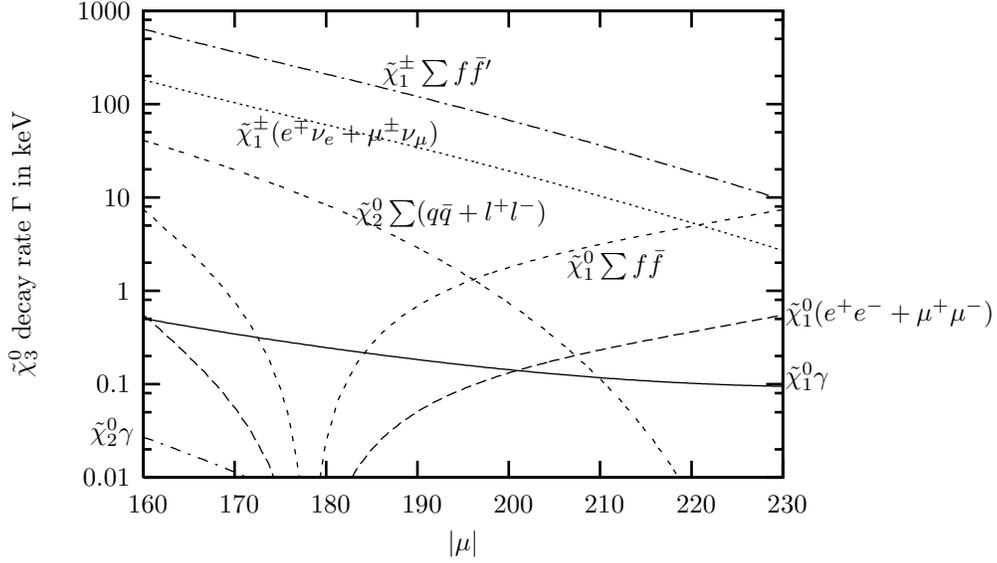}
  \caption{\label{fig:Gam3}
Partial decay widths of $\neu_3$
for $M_1=200$ GeV and $\tan\beta=10$.
All the CP violating phases are set to zero.
}
\end{figure}

\begin{figure}
\centering
 \includegraphics[scale=1.]{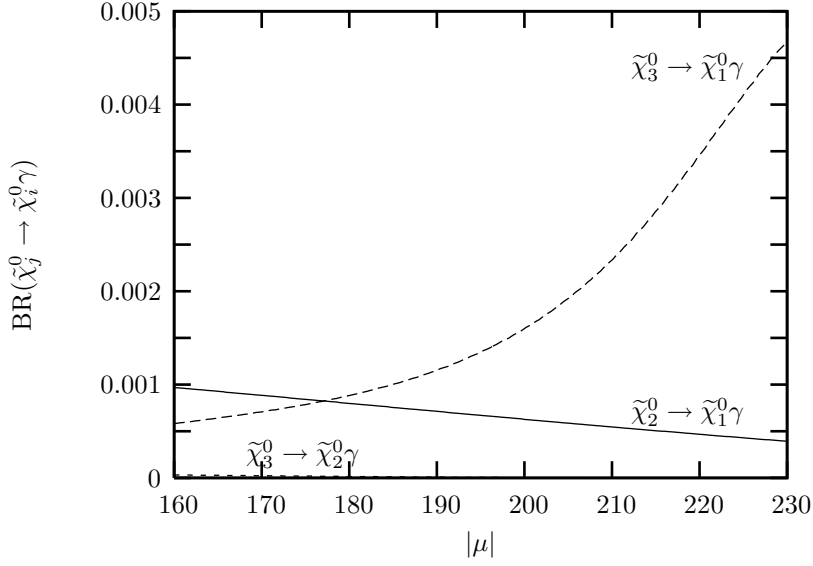}
  \caption{\label{fig:BRr}
Radiative decay branching ratios for $\neu_2$ and $\neu_3$
with $M_1=200$ GeV and $\tan\beta=10$.
All the CP violating phases are set to zero.
}
\end{figure}

First, we compute the partial decay widths of
$\neu_2$ and $\neu_3$ in the CP-conserving case
($\Phi_1 = \Phi_\mu =0$) in Figs. \ref{fig:Gam2}
and \ref{fig:Gam3}, respectively.
Considering the trigger strategy at the LHC,
we separate the electron and muon signal
from other fermion signals.
The notation of
$\sum f \bar f^{(')}$ includes
all appropriate fermions $f,f'$ except for $e^\pm$ and $\mu^\pm$.
In the
decay of $\neu_2$, the majority comes from
$\neu_2 \to \neu_1 f \bar f$ via an on-shell or
a virtual $Z$.  It is followed by
$\neu_2 \to \widetilde{\chi}^\pm f \bar f'$ via
an on-shell or a virtual $W$.
The radiative decay width is down by 3 orders of
magnitude from $\neu_2 \to \neu_1 Z^{(*)} \to \neu_1 f \bar f$
in the range of $\mu$ that we are interested in.
Figure \ref{fig:Gam3} shows the partial widths of
$\neu_3$.
A substantial mass difference between $m_{\neu_3}$ and $m_{\cham_1}$
for $|\mu|$ compatible with $M_1$
leads to the dominant decay mode into
$\cha f \bar{f}'$ via an on-shell or a virtual $W$.
It is followed by $\neu_3 \to \neu_2 Z^{(*)} \to \neu_2 f \bar f$,
which drops off quickly when $\mu$ increases.
A special dip (around $|\mu|\simeq 178$ GeV with $M_1 =200$ GeV)
shows up in the decay width of $\neu_3 \to \neu_1 Z^{(*)} \to
\neu_1 f \bar f$.  This is because of a cancellation
in the factor $(N_{13} N^*_{33} - N_{14} N^*_{34})$ of
the $\neu_3$-$\neu_1$-$Z$ coupling.
The radiative decay width of $\neu_3 \to \neu_1 \gamma$ is
relatively much larger than in the case of $\neu_2$.
This is the mode that we want to make use of in the
collider study of $\neu_3$ in split SUSY.
Since the masses of
$\neu_3$ and $\neu_1$ cannot be degenerate,
the outgoing photon is quite energetic.
A single-photon
plus missing energy event search is possible provided that
the branching ratio is large enough.
We show the radiative decay branching ratios for
$\neu_2 \to \neu_1 \gm$, $\neu_3 \to \neu_1 \gm$,
and $\neu_3 \to \neu_2 \gm$  in Fig. \ref{fig:BRr}.
The branching ratio for $\neu_3 \to \neu_2 \gamma$ is much smaller
than the other two.
With increasing $|\mu|$,
BR$(\neu_2 \to \neu_1 \gm)$ moderately decreases
while BR$(\neu_3 \to \neu_1 \gm)$ increases rather rapidly.

\begin{center}
\begin{figure}
  \includegraphics[scale=1.]{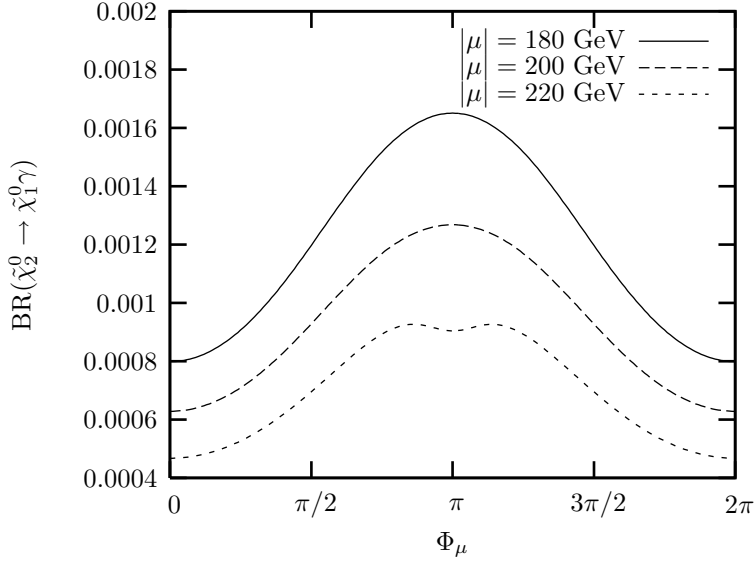}
  \caption{\label{fig:CPBRr21}
Radiative decay branching ratio of $\neu_2 \to\neu_1 \gamma$ as a function of
$\Phi_\mu$ (in unit of $\pi$) for
$M_1=200$ GeV and $\tan\beta=10$.
Three cases for $|\mu|=180,~200,~220$ GeV are shown.}
\end{figure}
\end{center}

\begin{center}
\begin{figure}
  \includegraphics[scale=1.]{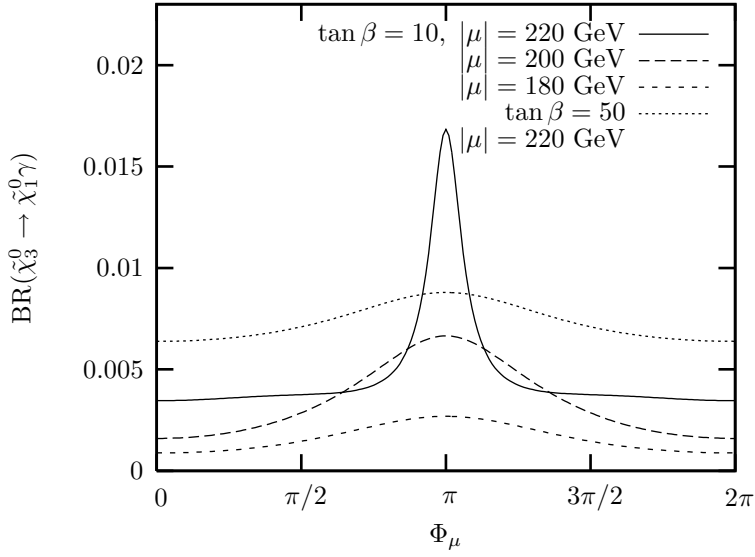}
  \caption{\label{fig:CPBRr31}
Radiative decay branching ratio of $\neu_3 \to\neu_1 \gamma$ as a function of
$\Phi_\mu$ (in unit of $\pi$) for
$M_1=200$ GeV and $\tan\beta=10$.
Three cases for $|\mu|=180,~200,~220$ GeV are shown.
One additional line (dotted) of $\tan\beta=50$ and $|\mu|=220$ GeV is also
shown.
}
\end{figure}
\end{center}

\begin{figure}
\centering
  \includegraphics[scale=1.]{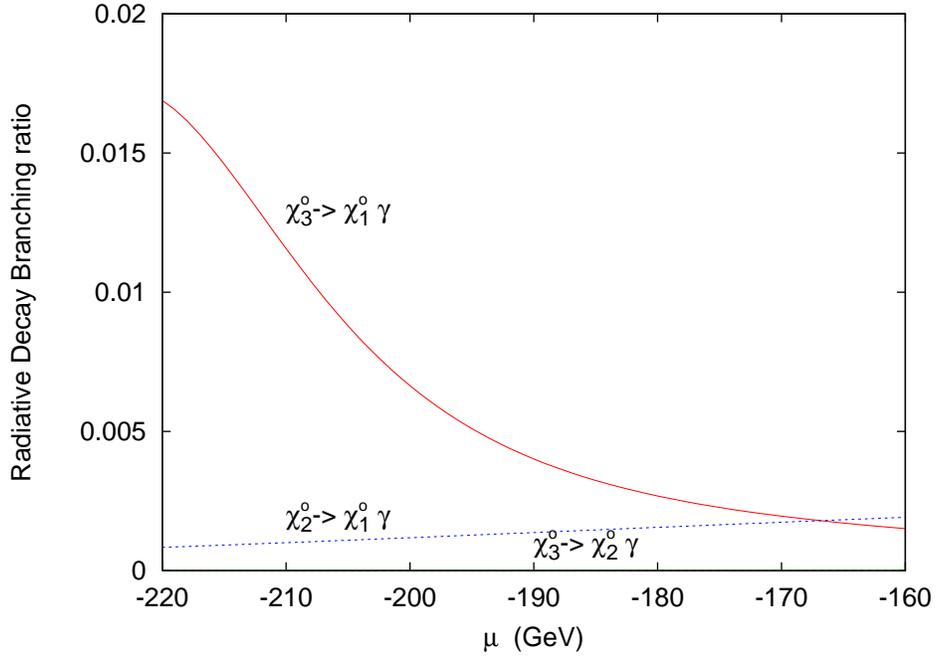}
  \caption{\label{fig:BRr-}
Radiative decay branching ratios for $\neu_2$ and $\neu_3$
with $M_1=200$ GeV and $\tan\beta=10$.
The $\Phi_\mu$ is set at $\pi$, i.e., negative $\mu$.  Note that
BR$(\widetilde{\chi}^0_3 \to \widetilde{\chi}^0_2 \gamma)$ is very close
to 0.
}
\end{figure}

Finally, we examine the dependence of CP-violating phases
on the neutralino radiative decay.
The effect of $\Phi_1$ is quite weak,
as expected, since $M_1$ is involved only in the neutralino mass matrix.
On the other hand,
the $\Phi_\mu$ dependence is rather strong.
The $\mu$ parameter affects both the neutralino and chargino sectors,
and the $\neu_j$-$\cha_i$-$W^\mp$ coupling in the loop.
We compute the $\Phi_\mu$-dependence on the radiative branching
ratios of $\neu_2$ and $\neu_3$ in Fig.~\ref{fig:CPBRr21} and
Fig.~\ref{fig:CPBRr31}, respectively.
Three cases of $|\mu|=180,~200,~220$ GeV are considered.
Figures \ref{fig:CPBRr21} and \ref{fig:CPBRr31}
clearly show the CP-phase sensitivity of the neutralino radiative
branching ratios.
As can be seen in Fig. \ref{fig:CPBRr31},
the BR$(\neu_3\to \neu_1 \gamma)$ can be enhanced
by a factor of about four with $|\mu|=220$ GeV at $\Phi_\mu = \pi$.
In most cases,
the maximum of radiative branching ratios
occurs at $\Phi_\mu=\pi$, i.e., negative $\mu$.
We also see the $\tan\beta$ dependence by plotting
BR$(\neu_3\to \neu_1 \gamma)$ for $\tan\beta=50$ and $|\mu|=220$GeV
in Fig. \ref{fig:CPBRr31}.
At $\Phi_\mu=0$, larger $\tan\beta$ enhances the radiative branching ratio
by a factor of about two.
However, at $\Phi_\mu=\pi$ where the radiative BR is maximized,
$\tan\beta=50$ case has smaller BR, about half of that with $\tan\beta=10$.
In Fig. \ref{fig:BRr-}, we show the variation of
the branching ratio versus negative $\mu$.  It is now clear that
at $\mu=-220$ GeV, a branching ratio about $1.68\%$ can be obtained
for the radiative decay $\neu_3 \to \neu_1 \gamma$ with $\tan\beta=10$.


\section{Production of neutralinos and charginos}
In split SUSY,
gaugino-pair production goes through Drell-Yan-like process
via $\gamma,Z$, or $W$ $s$-channel exchange diagrams.
Specifically, we study the production of
\beq
\label{eq:channel}
q + \overline{{q}} \rightarrow
\tilde{\chi}^0_i+\tilde{\chi}^0_j,\quad
q + \overline{{q}} \rightarrow
\tilde{\chi}^-_i+\tilde{\chi}^+_j,\quad
q + \overline{{q}}' \rightarrow
\tilde{\chi}^\pm_i+\neu_j\,.
\eeq
For simplicity we introduce the following notation:
\beq
\mu_{i\hs} = \frac{m_i^2}{\hs},\quad
D_X=\frac{\hs}{\hs-m^2_X+im_X\Gamma_X}\,,
\eeq
where $m_i$ denotes the mass of $\neu_i$ or $\cham_i$
and $X=Z,W$.

The helicity amplitude for neutralino-pair production is
\begin{eqnarray}
T\left(q \overline{q}\rightarrow\widetilde{\chi}^0_i\widetilde{\chi}^0_j\right)
 = \frac{e^2}{\hs}\sum_{\alpha,\beta=L,R}
 \mathcal{Q}^{(\mathrm{nn})ij}_{~\alpha\beta}
   \left[\bar{v}(\overline{q})  \gamma_\mu P_\alpha  u(q)\right]
   \big[\bar{u}(\widetilde{\chi}^0_i) \gamma^\mu P_\beta
               v(\widetilde{\chi}^0_j)\big]
               \,,
\label{eq:neutralino production amplitude}
\end{eqnarray}
where $\alpha,\beta=L,R$ and (nn) in the superscript of
$\mathcal{Q}^{(\mathrm{nn})ij}_{~\alpha\beta}$ denote neutralino
pair production. The four generalized bilinear charges
$\mathcal{Q}^{(\mathrm{nn})ij}_{~\alpha\beta}$ are
\beq
\mathcal{Q}^{(\mathrm{nn})ij}_{~\alpha\beta} = \frac{D_Z}{s_W^2
c_W^2}\, g^q_\alpha \,Q^{\rm nnZ}_{\beta ij} \,.
\eeq
Here the chiral
couplings of the quark $q$ with the $Z$ boson are given by
\beq g^q_R =
-s_W^2 Q_q,\quad g^q_L = (T^q_3)_L- s_W^2 Q_q \,,
\eeq
where $(T^q_3)_L$ is the third component of the isospin and
$Q_q$
is the electric charge of the quark $q$. 
The helicity amplitude for chargino pair production is
\begin{eqnarray}
T\left(q \overline{q}\rightarrow\widetilde{\chi}^-_i\widetilde{\chi}^+_j\right)
 = \frac{e^2}{\hs}\, \sum_{\alpha,\beta}
 \mathcal{Q}^{(\mathrm{cc})ij}_{~\alpha\beta}
   \left[\bar{v}(\overline{q})  \gamma_\mu P_\alpha  u(q)\right]
   \big[\bar{u}(\widetilde{\chi}^-_i) \gamma^\mu P_\beta
               v(\widetilde{\chi}^+_j)\big]
               \,,
\label{eq:chargino production amplitude}
\end{eqnarray}
where the bilinear charges are given by
\beq
\mathcal{Q}^{(\mathrm{cc})ij}_{~\alpha\beta}
=
-Q^q \delta_{ij}
+ \frac{D_Z}{s_W^2 c_W^2} g^q_\alpha Q^{\rm ccZ}_{\beta ij}
\,.
\eeq
Finally, the helicity amplitude for chargino-neutralino associated production
is
\begin{eqnarray}
T\left(d \bar{u}\rightarrow\widetilde{\chi}^-_i\neu_j\right)
 = \frac{e^2}{\hs}\, \sum_{\alpha,\beta}
  \mathcal{Q}^{(\mathrm{cn})ij}_{~\alpha\beta}
   \left[\bar{v}(\bar{u})  \gamma_\mu P_\alpha  u(d)\right]
   \big[\bar{u}(\widetilde{\chi}^-_i) \gamma^\mu P_\beta
               v(\widetilde{\chi}^0_j)\big]
               \,.
\label{eq:cha-neu}
\end{eqnarray}
where the bilinear charges are
\beq
\mathcal{Q}^{(\mathrm{cn})ij}_{~\alpha\beta} =
\frac{D_W}{\sqrt{2} s_W^2 } \,\delta_{\alpha L}\,Q^{\rm cnW}_{\beta ij}
\,.
\eeq

At parton level, the differential cross sections in the
 center-of-mass (c.m.) frame
for the above three channels have a common expression
\bea
\frac{{\rm d}\sigma}{{\rm d}\cos\theta^*}
  &=& \frac{1}{N_c}\frac{1}{\mathcal{S}}
  \frac{\pi\alpha^2}{2\,\hs}\, \lm^{1/2}\\ \no &&\times
  \Big[\left\{1-(\mu_{i\hs} - \mu_{j\hs})^2
                   +\lambda\cos^2\theta^*\right\}\mathcal{Q}_1^{\,ij}
                   +4\sqrt{\mu_{i\hs}\mu_{j\hs}}
 \mathcal{Q}_2^{\,ij}+2\lambda^{1/2} \mathcal{Q}_3^{\,ij}
                   \cos\theta^*\Big] \;,
\eea
where
$\theta^*$ is the scattering angle in the parton rest frame,
$\lm \equiv \lm(1,\mu_{i\hs},\mu_{j\hs})$,
$N_C$ is the color factor of $q$,
$\mathcal{S}$ is the symmetric factor
($\mathcal{S}=2$ for $\neu_i\neu_i$ production)
and $\theta^*$ is the scattering angle in the c.m. frame.
The $\mathcal{Q}_{1,2,3}^{\,ij}$ are combinations of bilinear charges
given by
\bea
\mathcal{Q}_1^{\,ij} &=&
\frac{1}{4}\left[|\mathcal{Q}_{RR}^{\,ij}|^2+|\mathcal{Q}_{LL}^{\,ij}|^2
              +|\mathcal{Q}_{RL}^{\,ij}|^2+|\mathcal{Q}_{LR}^{\,ij}|^2\right],
                       \\ \no
\mathcal{Q}_2^{\,ij}  &=&
\frac{1}{2}\;\real\left[\mathcal{Q}_{RR}^{\,ij}\mathcal{Q}_{RL}^{\,ij*}
                       +\mathcal{Q}_{LL}^{\,ij}\mathcal{Q}_{LR}^{\,ij*}\right],
                       \\ \no
\mathcal{Q}_3^{\,ij}  &=&
\frac{1}{4}\left[|\mathcal{Q}_{RR}^{\,ij}|^2+|\mathcal{Q}_{LL}^{\,ij}|^2
-|\mathcal{Q}_{RL}^{\,ij}|^2-|\mathcal{Q}_{LR}^{\,ij}|^2\right],
\eea
where the production channel of (nn), (cc) or (cn)
in the superscript is omitted.

\begin{figure}[t!]
\centering
\includegraphics{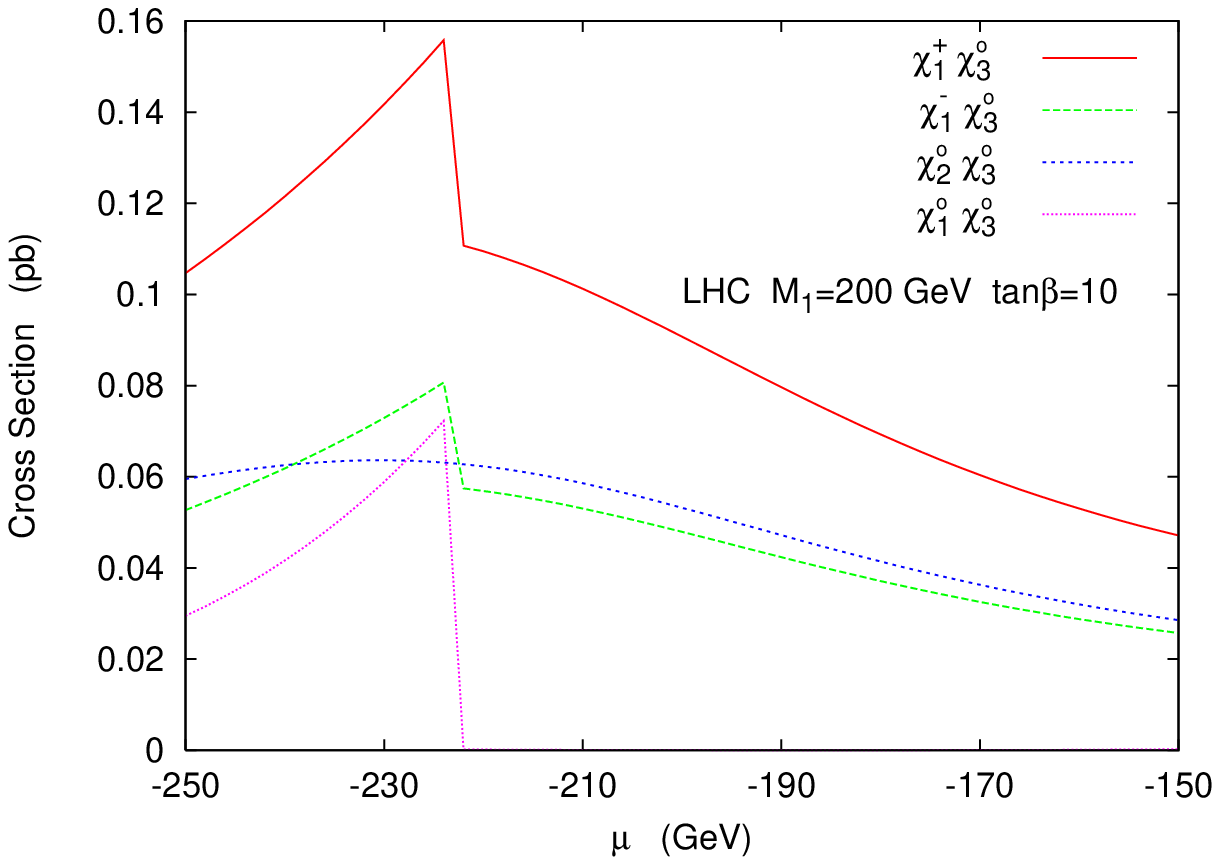}
\includegraphics{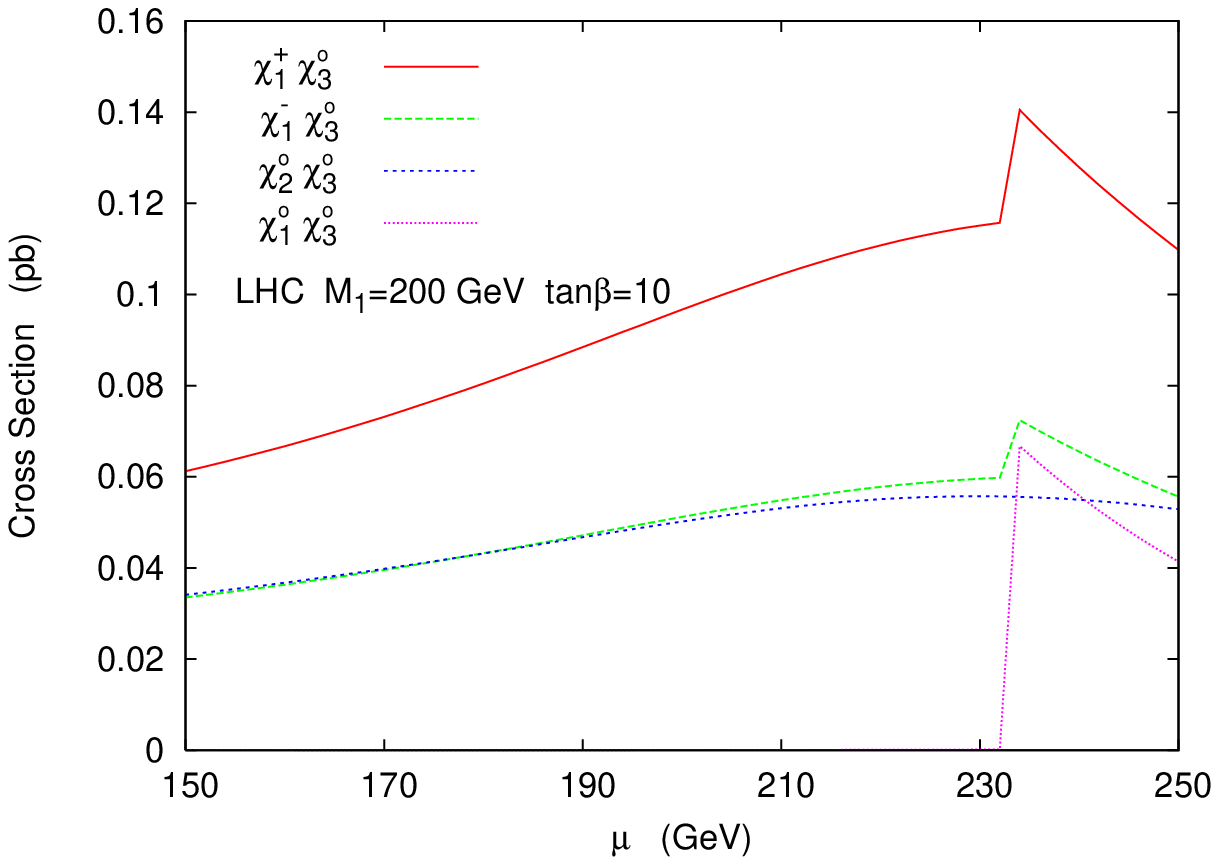}
\caption{\small \label{xs}
Production cross sections for $\neu_3$ at the LHC
versus (a) negative and (b) positive
$\mu$ with $M_1=200$ GeV and $\tan\beta=10$.
}
\end{figure}

In Fig. \ref{xs},
we show the production cross sections for $\neu_3$ at the LHC
versus negative and positive $\mu$ with $|M_1|=200$ GeV
and $\tan\beta=10$, including
$\widetilde{\chi}^0_3 \widetilde{\chi}^0_1$,
$\widetilde{\chi}^0_3 \widetilde{\chi}^0_2$,
$\widetilde{\chi}^0_3 \widetilde{\chi}^+_1$, and
$\widetilde{\chi}^0_3 \widetilde{\chi}^-_1$.
Here we consider only the $CP$-conserving case ($\Phi_\mu=\pi,0$).
It is clear  that the cross section is in general larger
in the region $|\mu| \sim M_1$, because this is region where the Bino and
Higgsino mix strongly.  This is also the favored parameter space for
the Bino-Higgsino dark matter\,\cite{giudice,aaron,stefano}.
Away from the peak the cross sections in general decrease as the mixing
between the Bino and the Higgsino becomes less; in particular, as
$|\mu|$ decreases from the mixing region ($M_1\sim\mu$) the
 $\widetilde{\chi}^0_1$ and $\widetilde{\chi}^0_2$ become Higgsino-like
while $\widetilde{\chi}^0_3$ becomes Bino-like, and so the
production of $\widetilde{\chi}^0_3$ with $\widetilde{\chi}^0_{1,2}$
and $\widetilde{\chi}^\pm_1$ all decrease rapidly.

The inclusive cross section for $\widetilde{\chi}^0_3$ is of the order of
$O(0.1-0.2)$ pb for $M_1=200$ GeV.  We have also calculated that
for $M_1=100$ GeV and maintaining the gaugino mass unification condition,
the inclusive cross section for $\neu_3$ is as large as $O(1-5)$ pb,
while the cross section goes down to about
$0.05$ pb for $M_1=300$ GeV.  Since we are interested in the radiative
decay of $\widetilde{\chi}^0_3$, which could have a branching ratio as
large as $O(1)$\%,
the number of single photon plus $\not \! E_T$ events
at the LHC is of the order of $O(10^3 - 5\cdot 10^3)$,
$O(100-200)$, and $O(50)$ for $M_1=100,200,300$
GeV, respectively.
The next-to-leading order (NLO) SUSY corrections to gaugino pair production
have been performed \cite{prospino}.  The $K$ factor is about
$1.1-1.4$ depending on SUSY parameters.  Since these SUSY corrections
would not affect significantly the photon momentum or missing energy
distributions, we just note that the event rates can be enhanced by up to
about 40\% due to NLO corrections.

\begin{center}
\begin{figure}
  \includegraphics[scale=1.]{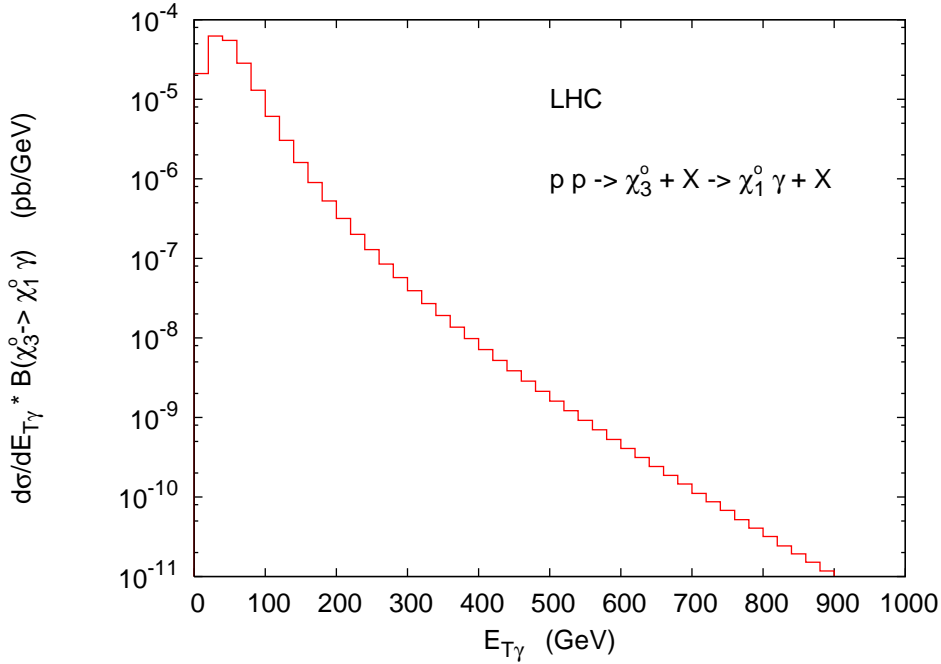}
  \caption{\label{fig:dsigdEtgamma}
Differential cross section versus the transverse momentum
of the photon for the process $pp \to \neu_3 +X \to\neu_1\gamma+X$. We set
$M_1=200$ GeV, $\mu = -220$ GeV, and BR$=1.68\%$.
}
\end{figure}
\end{center}

\begin{center}
\begin{figure}
  \includegraphics[scale=1.]{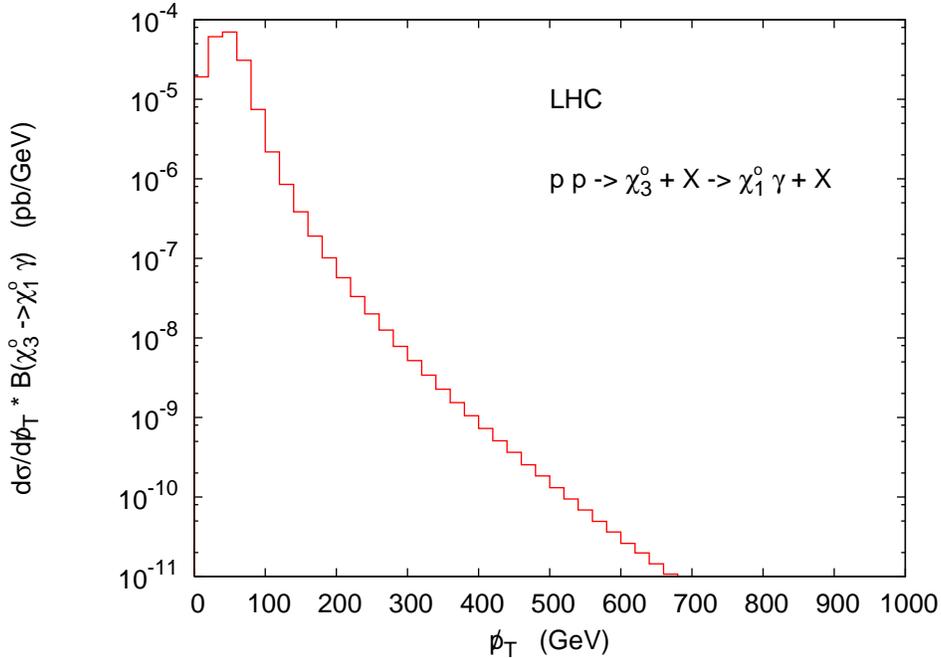}
  \caption{\label{fig:dsigdEmissing}
Differential cross section versus the missing transverse momentum
for the process $pp \to \neu_3 +X \to \neu_1 \gamma +  X$. We set
$M_1=200$ GeV, $\mu = -220$ GeV, and BR$=1.68\%$.
}
\end{figure}
\end{center}

Here we present the differential cross section versus the transverse
momentum of the single-photon and versus the missing transverse momentum.
We adopt a simple two-body decay of the third neutralino into the lightest
neutralino and the photon, without taking into account the spin correlation,
which should only be a mild effect.  We also assume the decay is prompt,
because the decay width of $\widetilde{\chi}^0_3$ is of the order of
MeV.  The contributing production channels include
$\widetilde{\chi}^0_3 \widetilde{\chi}^0_2$,
$\widetilde{\chi}^0_3 \widetilde{\chi}^0_1$, and
$\widetilde{\chi}^0_3 \widetilde{\chi}^\pm_1$,
We include all of these production channels to account for inclusive
$\widetilde{\chi}^0_3$ production.
We focus on the radiative decay of $\widetilde{\chi}^0_3
\to \widetilde{\chi}^0_1 \gamma$, which has
a branching ratio of the order of 1\% in the region $|\mu| \sim M_1$.
The charginos can decay into the neutralino and a virtual $W$ boson, which
further goes into $q\bar q'$ or $\ell \nu_\ell$.  Therefore, the final
state consists of an isolated single-photon with or without charged leptons or
jets, plus missing energies.  We are primarily interested in the
photon and missing energy distributions, which are shown in
Fig. \ref{fig:dsigdEtgamma} and Fig. \ref{fig:dsigdEmissing}, respectively.

Here we give a brief discussion on possible backgrounds.  An irreducible
background comes from the SM production of $\gamma Z$ followed by $Z\to
\nu\bar \nu$, as well as reducible background coming from quark
fragmentation into a photon.  Photon isolation and large $\not\!{p}_T$
cuts should be able to suppress the quark fragmentation background.
The $\gamma Z $ background, on average, has a smaller missing energy than
the signal, because very often in the signal there are two missing
neutralino LSP's.  Furthermore, a large portion of the signal will have
a charged lepton or jets coming from the associated chargino decay
that one can tag so as to remove the $\gamma Z$
background.
Therefore, with the above mentioned cuts, tagging, and isolations, the
signal of an isolated single-photon plus missing energy signature should be
rather clean.


\section{Conclusions}
We note that the single-photon plus missing energy signal 
is a possible sign of split SUSY, but not a unique feature.  For example,
gauge-mediated SUSY breaking (GMSB) models often predict various
signals consisting of single- or multi-photon plus multiple leptons 
and jets, and missing energy.
Split SUSY has to be established by gluino production, decay, and 
detection, and also with neutralino and chargino production and decay.
One has to determine the sfermion mass scale and the $\mu$ parameter
by simultaneously measuring the production and the decay properties of 
gauginos.
One can use the mass spectrum and the photon energy spectrum to 
differentiate between split SUSY and GMSB models.
We have shown that in general the 
$\widetilde{\chi}^0_3 \to \widetilde{\chi}^0_1
\gamma$ has a branching ratio of the order of 1\%, and the photon 
energy spectrum depends on the mass difference.
On the other hand, in the GMSB models when the NLSP is a neutralino, the
branching ratio into a photon and a gravitino is dominant.  Therefore,
the frequency of multi-photon plus missing energy events is very high.
Also, the photon energy spectrum depends on the mass of the 
lightest neutralino only.

In summary, 
we have studied the decays and productions of neutralinos and
charginos at hadron colliders in the scenario of split SUSY.
The decays of neutralinos are particularly interesting because
the sfermions are so heavy that only the decays via $Z^{(*)}$,
$W^{(*)}$, and $h^{(*)}$ are possible, among which the
radiative decay could have a branching ratio as large as $O(1)$\%,
unlike the case of MSSM.
We have found that $\neu_3 \to \neu_1 \gamma$ has a sufficiently
large branching ratio when $|\mu| \sim M_1$
such that isolated single-photon plus missing energy
events may be accessible in the LHC experiments.  We argued that
such events are rather clean and would be signs of split supersymmetry.
Furthermore, counting the event rates will also give a measurement on the
radiative decay branching ratio so that it gives information about
the parameters of split SUSY.

\begin{acknowledgments}
We appreciate the hospitality of KIAS as we completed a part of this
paper while we visit KIAS.
The work of KC was supported in part by
the National Science Council of Taiwan R.O.C. under grant no.
NSC 93-2112-M-007-025-.
The work of J. Song is supported by KRF under grant no. R04-2004-000-10164-0.
\end{acknowledgments}

\appendix
\section{Couplings of neutralinos and charginos}
All the couplings of supersymmetric fermions, neutralinos and charginos, to
gauge bosons or Higgs bosons are expressed by $Q^{ f f' X}_{\alpha i j}$.
Here $f(f')=\mathrm{c}$ in the superscript denotes the chargino,
$f(f')=\mathrm{n}$ denotes the neutralino,
$X=Z^\mu, W^\mu,h^0,H^\pm, G^\pm$,
and $\alpha=L,R$ in the subscript is the chirality of the fermions.
The $G^\pm$ is the Goldstone boson absorbed by $W^\pm$ boson.
In our notation, negatively charged chargino is assigned to be a particle
such that its electric charge $Q_{\cha}$ is $-1$.

The specific expressions for the $\cham_i$-$\chap_j$-$Z$ couplings are
\begin{eqnarray}
Q^{ {\rm ccZ}}_{L11} &=&
  s_W^2 - \frac{3}{4}-\frac{1}{4}
     \cos 2\phi_L \,,\qquad
Q^{ {\rm ccZ}}_{R11}
  = s_W^2-\frac{3}{4}-\frac{1}{4}\cos
    2\phi_R, \nonumber\\
Q^{ {\rm ccZ}}_{L12}
 &=&- \frac{1}{4}{\rm e}^{-i\beta_L}\sin 2\phi_L,\qquad\qquad
 ~
  Q^{ {\rm ccZ}}_{R12}
  = - \frac{1}{4} {\rm e}^{-i(\beta_R-\gamma_1+\gamma_2)}
                        \sin 2\phi_R,\nonumber\\
Q^{ {\rm ccZ}}_{L22}
  &=&  s_W^2 - \frac{3}{4}+\frac{1}{4}
     \cos 2\phi_L \,,\qquad
Q^{ {\rm ccZ}}_{R22}
 =  s_W^2-\frac{3}{4}+\frac{1}{4}\cos
    2\phi_R .\nonumber
\end{eqnarray}
The $\neu_i$-$\neu_j$-$Z$ couplings are
\beq
 Q^{ \rm nnZ}_{L ij}
  = \frac{N_{i3}N^*_{j3}-N_{i4}N^*_{j4}}{2}
\,,
\eeq
where $Q^{ \rm nnZ}_{R ij} =-(Q^{ \rm nnZ}_{L ij})^*$.
Neutral couplings of $\neu_i$-$\neu_j$-$h^0$,
where $h^0$ is the light $CP$-even Higgs boson,
are divided into $i=j$ and $i \neq j$ cases, by using Majorana conditions:
\bea
S_i^{\rm nnh} &=&
\frac{1}{2}
\left[
N_{i2}-t_W N_{i1}
\right]
\left[
-\sin\alpha N_{i3}-\cos\alpha N_{i4}
\right],\\
Q^{\rm nnh}_{Lij} &=&\frac{1}{2}
\left[(N_{i2}^* - t_W N^*_{i1})
(- \sin\alpha N^*_{j3} -\cos\alpha N^*_{j4})
+(i \leftrightarrow j)
\right]
\,,
\eea
where $Q^{\rm nnh}_{Rij}=(Q^{\rm nnh}_{Lij})^*$, $t_W=\tan\theta_W$ and
\beq
\tan\alpha =
\frac{m_h^2-m_A^2\cos^2\beta-m_Z^2\sin^2\beta}
{(m_A^2+m_Z^2)\sin\beta\cos\beta}
\,.
\eeq

The  $\cham_i$-$\neu_j$-$W^+$ couplings are
\beq
Q^{\rm cnW}_{Lij}=
(U_L)_{i1}N_{j2}^*
+\frac{1}{\sqrt{2}}
(U_L)_{i2}N^*_{j3}
 ,
\quad
Q^{\rm cnW}_{Rij}=
(U_R)_{i1}N_{j2}
-\frac{1}{\sqrt{2}}
(U_R)_{i2}N_{j4}
\,.
\eeq
The charged couplings to $H^\pm$ are
\bea
Q^{\rm cnH}_{Lkj} &=&
\cos\beta
\left[
(U_R)_{k1}N_{j4}^*+\frac{(U_R)_{k2}}{\sqrt{2}}(N_{j2}^*+t_W N^*_{j1})
\right],
\\
\nonumber
Q^{\rm cnH}_{Rkj} &=&
\sin\beta
\left[
(U_L)_{k1}N_{j3}-\frac{(U_L)_{k2}}{\sqrt{2}}(N_{j2}+t_W N_{j1})
\right].
\eea
Finally we have the $\cham_i$-$\neu_j$-$G^+$ couplings of
\beq
Q^{\rm cnG}_{\alpha kj} =Q^{\rm cnh}_{\alpha kj}
\left(
\sin\beta \to -\cos\beta,~\cos\beta \to \sin\beta
\right)\,.
\eeq

\end{document}